\documentclass[preprint,superscriptaddress,nofootinbib]{revtex4}
\usepackage{amsmath}
\usepackage{amssymb}
\usepackage{graphicx}
\usepackage{mathrsfs}
\usepackage{cancel}
\usepackage{soul}
\usepackage{color}
\usepackage{multirow}
\usepackage{bm}
\usepackage[utf8]{inputenc}

\begin{document}

\title{An optimal scheme for top quark mass measurement near $t\bar{t}$ threshold at future $e^{+}e^{-}$ colliders}
\author{Wei-Guo Chen}
\email{chenwg19910708@snnu.edu.cn}
\affiliation{School of Physics $\&$ Information Technology, Shaanxi Normal University, Xi'an 710119, China}
\author{Xia Wan}
\email{wanxia@snnu.edu.cn}
\affiliation{School of Physics $\&$ Information Technology, Shaanxi Normal University, Xi'an 710119, China}
\author{You-Kai Wang}
\email{wangyk@snnu.edu.cn}
\affiliation{School of Physics $\&$ Information Technology, Shaanxi Normal University, Xi'an 710119, China}

\date{\today}

\begin{abstract}
 A simulation of top quark mass measurement scheme near the $t\bar{t}$ production threshold in future $e^{+}e^{-}$ colliders, e.g.\,the Circular Electron Positron Collider(CEPC), is performed. $\chi^2$ fitting method is adopted to determine the number of energy points to be taken and their locations. Our result shows that the optimal energy point is located near the largest slope of the cross section to beam energy and the most efficient scheme is to concentrate all luminosity on this single energy point in one parameter top mass fitting case. This suggests that the so called data driven method can be a best choice for the future real experimental measurement. Conveniently, the top mass statistical uncertainty can also be calculated directly by the error matrix even without any sampling and fitting. Agreement of the above two optimization methods has been checked. Our conclusion is that by taking 50~$fb^{-1}$ total effective integrated luminosity data, the statistical uncertainty of the top potential subtracted mass can be suppressed to about 7~MeV and the total uncertainty is about 30~MeV. This precision will help to identify the stability of the electroweak vacuum at the Planck scale.
\end{abstract}

\maketitle

\section{Introduction}
The Higgs potential is closely related to both the Higgs boson mass and the top quark pole mass. Especially, if the top quark mass is too heavy, the quartic Higgs coupling $\lambda$ in the Standard Model may be negative at large energy scale before the Planck scale and the stability of electroweak vacuum breaks. Therefore, the determination of the electroweak vacuum stability needs precise measurements for both the Higgs boson mass and the top quark mass. At the Large Hadron Collider (LHC), the Higgs mass is measured with the precision of $\mathcal{O}$(200)~MeV~\cite{pdg}, which means, currently, the electroweak vacuum stability is more sensitive to the uncertainty of the top quark pole mass.

Before detailed investigation of the top quark mass, one should keep in mind that the top quark mass is not an experimental direct observable. This means the value of the experimental output masses should depend on the theoretical input definitions.

Theoretically, kinds of top masses can be defined.
\begin{itemize}
\item Pole mass

The pole mass has an inherent ambiguity in order of $\mathcal{O}(\Lambda_{\rm{QCD}})$~\cite{mpoleuncertain,mpole_ambiguity,ambiguity} and leads to an instability of top threshold peak location at different orders, thus the pole mass is not a good definition for experimental measurements and unambiguous definitions of top masses are necessary. The renormalized top quark propagator is expressed as
\begin{equation}
D( /\kern-0.58emp) = \dfrac{i}{/\kern-0.58em p - m_{R} - \sum_R(/\kern-0.58em p)}.
\end{equation}
From the denominator, we have

\begin{equation}
\label{demoninator}
/\kern-0.58emp_{pole} = m_{R} + \mbox{$\sum_R$}(/\kern-0.58em p),
\end{equation}

where $m_R$ is the renormalized top mass, $\mbox{$\sum_R$}(/\kern-0.58em p)$ is the renormalized top quark self-energy contribution. At $\alpha_s$ first order of the top quark self-energy, it can be expressed as

\begin{equation}
\label{selfenergy}
\mbox{$\sum_{R}^{(1)}$}(/\kern-0.58em p)=m_{R}\sum^{\infty}_{n=0}c_{n}\alpha_{s}^{n+1}(m_R),
\end{equation}
where the coefficient $c_n\rightarrow 2^nn!$ and the convergence of perturbative expansion breaks when $n\rightarrow \infty$. The behavior of this IR renormalon results an intrinsic ambiguity of the pole mass. The ambiguity is estimated as~\cite{ambiguity}
\begin{equation}
%\begin{aligned}
\delta m_{pole} = \dfrac{C_{F}}{2N_{f}\vert \beta _{0} \vert}e^{-C/2}\Lambda _{QCD} \left( \ln\dfrac{m^{2}}{\Lambda ^{2}_{QCD}} \right)^{\beta_1/(2 \beta ^{2}_{0})}\sim \Lambda_{QCD},
%\end{aligned}
\end{equation}
where $\beta_i$ is the $i+1$th-loop beta function, $C$ is a constant related to renormalization scheme,
%($C=-5/3$ in $\overline{\mbox{MS}}$ scheme),
$C_F=4/3$.
\end{itemize}

To avoid the pole mass ambiguity, several short distance masses can be defined due to the IR sensitive term cancellation between the pole mass and the static potential $V(r)$ of the toponium.

\begin{itemize}
\item Potential subtracted(PS) mass

From the conservation of the total energy, we have
\begin{equation}
2m_{pole}+V(r) = 2m_{PS}+V(r,\mu_f),
\end{equation}
where $V(r,\mu_f)$ is the subtracted potential and can be defined as~\cite{mPS}
\begin{equation}
V(r,\mu_f)=V(r)-\int_{|\vec{q}|<\mu_{f}}\dfrac{d^{3}\vec{q}}{(2\pi)^{3}}\tilde{V}(\vec{q}).
\end{equation}
At $\alpha_s$ leading order, $\tilde{V}(\vec{q})=-\dfrac{4\pi C_F\alpha_s(\mu)}{\vec{q}^2}$ is the potential in momentum space. So the relations between difference masses are
\begin{equation}
\begin{aligned}
m_{PS} &= \dfrac{1}{2}[2m_{pole} + V(r) - V(r,\mu_{f})] \\
&= m_{pole} + \dfrac{1}{2}\int_{|\vec{q}|<\mu_{f}}\dfrac{d^{3}\vec{q}}{(2\pi)^{3}}\tilde{V}(\vec{q}).
\end{aligned}
\end{equation}
By considering Eq.(\ref{demoninator})and (\ref{selfenergy}),  we have
\begin{equation}
\begin{aligned}
m_{PS} &= m_{R}(1+\sum^{\infty}_{n=0}c_{n}\alpha_{s}^{n+1}(m_R)) - \dfrac{1}{2}\int_{|\vec{q}|<\mu_{f}}\dfrac{d^{3}\vec{q}}{(2\pi)^{3}}\dfrac{4\pi C_F \alpha_s(\mu)}{q^2}\\
&=m_{R}(1+\sum^{\infty}_{n=0}c_{n}\alpha_{s}^{n+1}(m_R))-\mu_f \sum^{\infty}_{n=0}c'_{n}\alpha_s ^{n+1}(m_R).
\end{aligned}
\end{equation}
We see that the coefficients $c_n$ and $c'_n$ should have the same divergent form($c_n$, $c'_n$ $\rightarrow 2^nn!$) as $n\rightarrow \infty$, thus the IR renormalons are cancelled exactly and only the non-ambiguous terms remain. It should be pointed out that the remained coefficient $\mu_f$ can not be removed. This is why the PS mass depends on the scale $\mu_f$ when it is expressed by other shot-distance masses(such as the $\overline{\mbox{MS}}$ mass).

\item 1S mass

The 1S mass is defined as half of the perturbative mass of the toponium $1\ ^3\mbox{S}_1$ ground state and is given by~\cite{1Smass,toppik}

\begin{equation}
m_{1S}=\dfrac{1}{2}\left( 2m_{pole}+E_{1S}(m_{pole},\alpha_s(\mu)) \right),
\end{equation}
where
\begin{equation}
E_{1S}(m_{pole},\alpha_s(\mu))=\int \dfrac{d^3\vec{p}}{(2\pi)^3} \dfrac{d^3\vec{q}}{(2\pi)^3}\tilde{\psi}^{*}_{1S}(\vec{p}) \tilde{H}(\vec{p},\vec{q}) \tilde{\psi}_{1S}(\vec{q}),
\end{equation}
$\tilde{H}(\vec{p},\vec{q})$ and  $\tilde{\psi}_{1S}$
are the Hamiltonian and the wave function in the $1\ ^3 S_1$ state in momentum space respectively.

Considering the IR behavior
\begin{equation}
\begin{aligned}
E^{IR}_{1S}(m_{pole},\alpha_s(\mu))&=\int_{IR} \dfrac{d^3\vec{p}}{(2\pi)^3} \dfrac{d^3\vec{q}}{(2\pi)^3}\tilde{\psi}^{*}_{1S}(\vec{p}) \tilde{H}(\vec{p},\vec{q}) \tilde{\psi}_{1S}(\vec{q})\\
&=\int_{IR} \dfrac{d^3\vec{p}}{(2\pi)^3} \dfrac{d^3\vec{q}}{(2\pi)^3}\tilde{\psi}^{*}_{1S}(\vec{p})\left(\dfrac{\vec{p}^2}{2m_{pole}}+\dfrac{\vec{q}^2}{2m_{pole}}+\tilde{V}(\vec{p}-\vec{q})\right) \tilde{\psi}_{1S}(\vec{q}).
\end{aligned}
\end{equation}
Dropping the momentum terms in Hamiltonian and denoting the IR region $\vert \vec{q} \vert,\vert \vec{p} \vert < \mu_f$,
\begin{equation}
\begin{aligned}
E^{IR}_{1S}(m_{pole},\alpha_s(\mu)) &\sim \int_{IR} \dfrac{d^3\vec{p}}{(2\pi)^3} \dfrac{d^3\vec{q}}{(2\pi)^3}\tilde{\psi}^{*}_{1S}(\vec{p})\tilde{V}(\vec{p}-\vec{q}) \tilde{\psi}_{1S}(\vec{q}) \\
&\sim \int_{\vert \vec{q} \vert,\vert \vec{p} \vert < \mu_f} \dfrac{d^3\vec{p}}{(2\pi)^3} \dfrac{d^3\vec{q}}{(2\pi)^3}\vert \tilde{\psi}^{*}_{1S}(\vec{p}) \vert^2\tilde{V}(\vec{p}-\vec{q})  \\
&\sim \int_{\vert \vec{q} \vert < \mu_f} \dfrac{d^3\vec{q}}{(2\pi)^3}\tilde{V}(\vec{q}).
\end{aligned}
\end{equation}

We see that it is very like the PS mass case as the IR behavior of the $E_{1S}(m_{pole},\alpha_s(\mu))$ results an IR renormalon which cancels with the ambiguity of $m_{pole}$. Thus the 1S mass contains non-ambiguity.

\item $\overline{\mbox{MS}}$ mass: defined by the modified minimal subtraction renormalization scheme.
\end{itemize}

Experimentally, the top quark mass can be measured mainly by two methods. The first one is from the top decay products reconstruction~\cite{method}. For example, the current most precise top mass is obtained from the lepton+jets channel. The main source of errors comes from the jet energy scale calibration.  However, the experimental measured top mass corresponds to none of the above theoretical mass definitions. The reason comes from the Monte Carlo(MC) simulation to determine the selection efficiency. As we need to put in an initial top mass to generate events in MC, the final measured top mass will inevitably be affected by this initial mass. So the experimental measured top mass is usually named as the MC mass. Approximately, sometimes people do not distinguish the MC mass and the pole mass as their difference is estimated to be less than 1~GeV.

The second method is extracting top mass from measured $t\bar{t}$ cross section by comparing it with the theoretical cross section~\cite{chi2,method1,Chatrchyan:2013haa,Abazov:2011pta}. The two cross section curves have different dependent relations on the fictional top mass and the overlap region corresponds to the real top mass. The advantage of this method is that it has a relatively clear mass definition (not absolutely clear as it also needs MC simulation), but the accuracy is not so good.

Current PDG values are~\cite{pdg}
$$
\begin{array}{ll}
\mbox{Direct measurement} & m=173.1\pm0.6\mbox{GeV},\\
\mbox{Mass from cross section measurements}  & m=160^{+5}_{-4}\mbox{GeV},\\
\mbox{Pole from cross section measurements} & m=173.5\pm1.1\mbox{GeV}.

\end{array}
$$

Alternatively, there is a third method which uses top pair threshold scan at future 350~GeV $e^{+}e^{-}$ colliders, e.g.\,the International Linear Collider(ILC), the Compact Linear Collider(CLIC), the $e^{+}e^{-}$ Future Circular Collider(FCC-ee) and CEPC etc.. The corresponding simulations have been performed in~\cite{simulTop,CLIC,N3LO1}. Because of the clear mass definition, sensitive dependence of the cross section on the top mass and low background pollution, this method is believed to be the best choice to obtain the most accurate top mass although it's expensive and time consuming.

This paper is organized as follows: In Sec.\,\uppercase\expandafter{\romannumeral2}, we review the framework of threshold top pair production cross section which is declared up to NNNLO QCD level.
\iffalse and show a numerical computation by \texttt{QQbar$\_$threshold} code~\cite{QQbar_threshold}\fi
In Sec.\,\uppercase\expandafter{\romannumeral3}, we perform Possion-Sampling and $\chi^{2}$ fitting by using \texttt{Minuit}~\cite{minuit} code, and present an equivalent error matrix analysis for the statistical uncertainty estimation. Our study shows that the most efficient data taking strategy is  just one optimal energy point which locates in the largest slope of the cross section to the beam energy region. The so-called  data-driven method is necessary. In Sec.\,\uppercase\expandafter{\romannumeral4}, we discuss briefly the impact of future CEPC top mass measurement on the electroweak vacuum stability. In Sec.\,\uppercase\expandafter{\romannumeral5}, we give a short summary.

\section{cross section}\label{section2}
The theoretical high order QCD calculations of the cross section \emph{$e^{+}$}\emph{$e^{-}$}$\to$$\gamma^{*}$/$Z^{*}$$\to$\emph{t$\bar{t}$} near threshold is built
in the framework of nonrelativistic quantum chromodynamics(NRQCD) \cite{NRQCD,NRQCD1} and potential nonrelativistic quantum chromodynamics(pNRQCD) \cite{PNRQCD}. NRQCD is obtained by integrating out the hard part $\mathcal{O}$(\emph{m}) of the QCD and pNRQCD is obtained by integrating out the soft part $\mathcal{O}$(\emph{m}$\upsilon$) of the NRQCD. The top pair production cross section at NNLO QCD order appeared in the 1990s, e.g.\,Ref.\,\cite{toppik}, and recently has been updated to NNNLO QCD \cite{N3LO}. On the other hand, when the energy approaches to the threshold, the top quark velocity $\upsilon$ becomes very small. The corresponding resummation for Coulomb singularities and large logarithms is completed at next-to-next-to-leading-logarithmic order(NNLL) in \cite{resum}. The details can been found in \cite{m1S1}. These are implemented in Monte Carlo generators \texttt{Whizard} \cite{whizard}, which could make multi-particle process simulations at $e^{+}e^{-}$ colliders. It includes a model `` SM$\_$tt$\_$threshold.mdl " that can be used to calculate the top pair production cross section near the threshold at LL order and NLL order \cite{NLL, NLL1, NLL2}. Because the top quark pair is unstable and decays to $W^{+}W^{-}b\bar{b}$ instantaneously when they are produced, the full process $e^{+}e^{-}$ $\rightarrow$ \emph{$W^{+}W^{-}b\bar{b}$} should be taken into account, thus it has backgrounds which come from the decay of $W^{+}W^{-}$, \emph{ZZ} and \emph{ZH} etc. As pointed out in Ref.\,\cite{phacut}, these backgrounds can increase the total cross section. In order to reduce these backgrounds, invariant mass cuts for \emph{$W^{+}$b} and \emph{$W^{-}\bar{b}$} are needed and can be taken the form $\mid\emph{M}_{W,b}-\emph{m}_{t}\mid \leq \bigtriangleup\emph{M}_{t}$. The analysis \cite{phacut1} shows that a cut with $\bigtriangleup\emph{M}_{t} \backsim$ 15 - 35 GeV is moderate, so in our calculations we set $\bigtriangleup\emph{M}_{t}$ = 30~GeV.

In the followings, we briefly review the theoretical framework of the total cross section calculations for top pair bound state. The top pair total cross section can be written in the form
\begin{equation}\label{totalcrosssection}
\begin{aligned}
\sigma(e^{+}e^{-} \rightarrow \emph{t$\overline{t}$} + \emph{X}) = \sigma_{0}\cdot(R^{\upsilon} + R^{a}),
\end{aligned}
\end{equation}
where $\sigma_{0}$ = 4$\pi\alpha^{2}$/3\emph{s} is the cross section for the $\mu^{+}\mu^{-}$ pair at tree level and $s= q^{2} =(E+2m_{t})^{2}$ is the square of center of mass energy, $R^{\upsilon}$ and $R^{a}$ are the ratios contributed by vector current and axial-vector current respectively, which can be related to the two-point functions of the vector current and the axial-vector current separately by the optical theorem,
\begin{equation}\label{ratio}
\begin{aligned}
&R^{\upsilon} = [(e_{t} - \dfrac{q^{2}\upsilon_{e}\upsilon_{t}}{q^{2} - m^{2}_{Z}})^{2} + (\dfrac{q^{2}}{q^{2}-m^{2}_{Z}})^{2}\cdot a^{2}_{e}\upsilon^{2}_{e}]\,\mathcal{I}m(\Pi^{\upsilon}(q^{2})),\\
&R^{a} = (\dfrac{q^{2}}{q^{2} - m^{2}_{Z}})^{2}(\upsilon^{2}_{e} + a^{2}_{e})a^{2}_{t}\,\mathcal{I}m(\Pi^{a}(q^{2})),
\end{aligned}
\end{equation}
where the vector and axial-vector couplings of fermions to the $Z$ boson are
\begin{equation}
\upsilon_{f} = \dfrac{T^{f}_{3} - 2e_{f}\sin^{2}\theta_{w}}{2\sin\theta_{w}\cos\theta_{w}},\quad a_{f} = \dfrac{T^{f}_{3}}{2\sin\theta_{w}\cos\theta_{w}},
\end{equation}
$e_{f}$ is the electric charge of the fermion in units of positron charge ($e_{f}$ = 2/3 for top quark and $e_{f}$ = 1 for electron), $T^{f}_{3}$ is the third component of its weak isospin and $\theta_{w}$ denotes the Weinberg angle.

The two-point Green function of vector (or axial-vector) current are given by \cite{two-point-function}
\begin{equation}\label{Green-func}
\begin{aligned}
\Pi^{X}_{\mu\nu} &= i \int d^{4}x e^{iq\cdot x}\langle0\vert Tj^{X}_{\mu}(x)j^{X}_{\nu}(0) \vert0\rangle \\
 &= (q_{\mu}q_{\nu} - q^{2}g_{\mu\nu})\Pi^{X}(q^{2}) + q_{\mu}q_{\nu}\Pi^{X}_{L}(q^{2}),
\end{aligned}
\end{equation}
where $X=v$ (or $X=a$) denotes the vector (or axial-vector) current, and $j^{X}_{\mu}$ = $\bar{t}\gamma_{\mu}t$ (or $\bar{t}\gamma_{\mu}\gamma_{5}t$).

In the framework of (p)\,NRQCD, the expansion of the vector and axial-vector currents read
\begin{equation}\label{current}
\begin{aligned}
j^{\upsilon}_{k} = c_{\upsilon}\psi^{\dagger}\sigma_{k}\chi + \dfrac{d_{v}}{6m^{2}_{t}}\psi^{\dagger}\sigma_{k}\textbf{D}^{2}\chi + ..., \quad j^{a}_{k} = \dfrac{c_{a}}{2m_{t}}\psi^{\dagger}[\sigma_{k},(-i)\vec{\sigma} \cdot\textbf{D}]\chi + ...,
\end{aligned}
\end{equation}
where $\sigma_k$ is Pauli matrix and $\textbf{D}=-\vec{\bigtriangledown}$, $\psi$ is the top quark field and $\chi$ is the anti-top quark field, $c_{v}$, $d_{v}$ and $c_{a}$ are the non-relativistic QCD\,(NRQCD) matching coefficients of vector and axial-vector currents\cite{N3LO_detail,two-point-function}. In the center of mass frame\,(CM frame), the momentum $q^{\mu}$ = ($q^{0},\textbf{0}$) = ($E+2m_{t},\textbf{0}$). From Eq.\,(\ref{Green-func}), one can easily rewrite Eq.\,(\ref{ratio}) in $d$-dimension space, and
\begin{equation}\label{ratio_detail}
\begin{aligned}
&R^{\upsilon} = \dfrac{1}{(d-1)q^{2}}[(e_{t} - \dfrac{q^{2}\upsilon_{e}\upsilon_{t}}{q^{2} - m^{2}_{Z}})^{2} + (\dfrac{q^{2}}{q^{2}-m^{2}_{Z}})^{2}\cdot a^{2}_{e}\upsilon^{2}_{e}]\,\mathcal{I}m(i \int d^{4}x e^{iq^{0}\cdot x^{0}}\langle0\vert Tj^{X}_{k}(x)j^{X}_{k}(0) \vert0\rangle),\\
&R^{a} = \dfrac{1}{(d-1)q^{2}}(\dfrac{q^{2}}{q^{2} - m^{2}_{Z}})^{2}(\upsilon^{2}_{e} + a^{2}_{e})a^{2}_{t}\,\mathcal{I}m(i \int d^{4}x e^{iq^{0}\cdot x^{0}}\langle0\vert Tj^{X}_{k}(x)j^{X}_{k}(0) \vert0\rangle).
\end{aligned}
\end{equation}
By inserting Eq.\,(\ref{current}) into Eq.\,(\ref{ratio_detail}) and making the fallowing substitutions
\begin{equation}\label{G(E)}
\begin{aligned}
&G^{S}(E) = \dfrac{i}{2N_{c}(d-1)}\int d^{4}x e^{iEx^{0}} \langle0\vert T(\chi^{\dagger}\sigma_{k}\psi)(x)(\psi^{\dagger}\sigma_{k}\chi)(0) \vert0\rangle,\\
&G^{P}(E) = \dfrac{i}{2N_{c}}\int d^{4}x e^{iEx^{0}} \langle0\vert T(\chi^{\dagger}iD_{k}\psi)(x)(\psi^{\dagger}iD_{k}\chi)(0) \vert0\rangle,
\end{aligned}
\end{equation}
where the superscripts ``$S$'' and ``$P$'' denote the $S$-wave state and the $P$-wave sate respectively.
The $R^{\upsilon}$ and $R^{a}$ therefore are simplified to
\begin{equation}\label{ratio_result}
\begin{aligned}
&R^{\upsilon} = [(e_{t} - \dfrac{q^{2}\upsilon_{e}\upsilon_{t}}{q^{2} - m^{2}_{Z}})^{2} + (\dfrac{q^{2}}{q^{2}-m^{2}_{Z}})^{2}\cdot a^{2}_{e}\upsilon^{2}_{e}]\cdot\dfrac{N_{c}}{2m^{2}_{t}}c_{v}[c_{v}-\dfrac{E}{m_{t}}(c_{v}+\dfrac{d_{v}}{3})]\mathcal{I}m\lbrace G^{S}(E)\rbrace,~\\
&R^{a} = (\dfrac{q^{2}}{q^{2} - m^{2}_{Z}})^{2}(\upsilon^{2}_{e} + a^{2}_{e})a^{2}_{t} \cdot\dfrac{N_{c}c^{2}_{a}}{2m^{4}_{t}}\dfrac{d-2}{d-1}\mathcal{I}m\lbrace G^{P}(E)\rbrace.
\end{aligned}
\end{equation}
In (p)NRQCD perturbation theory, the expansion for the Green function $G^{S}(q^{0})$ takes
\begin{equation}\label{expansion of G(E)}
\begin{aligned}
G^{X}(E) = G^{X}_{0}(E) + \sum^{n}_{i=1}\delta_{i}G^{X}(E),
\end{aligned}
\end{equation}
where $G^{X}_{0}(E) = G^{X}_{0}(0,0;E)$ is the zero-point Green function in coordinate space, which can be derived by solving the non-relativistic Schr\"odinger equation in spherical coordinate
\begin{equation}\label{schrodinger equation}
\left[-\dfrac{1}{m_{t}}\left(\dfrac{d^{2}}{dr^{2}}+\dfrac{2}{r} \dfrac{d}{dr}\right)-\dfrac{C_{F}\alpha_{s}}{r}-E \right]G_{0}(\textbf{r},\textbf{r}^{'};E)=\dfrac{1}{4\pi r^{2}}\delta(\textbf{r}-\textbf{r}^{'}),
\end{equation}
 $\delta_{i}G^{X}(E)$(i = 1, 2, 3, ...) are the high order corrections, which are not only related to the high order corrections of Coulomb potential but also $G^{X}_{0}(E)$. The complete $S$-wave $G^{X}(E) = G^{S}(E)$ calculations at third order are provided in Ref.\,\cite{N3LO_detail} and the $P$-wave $G^{X}(E) = G^{P}(E)$ at the same order is included in Ref.\,\cite{two-point-function}. The complete numerical calculations are implemented in the \texttt{QQbar$\_$threshold} code \cite{QQbar_threshold}.

\begin{figure}[htbp]
\centering
\centerline{\includegraphics[height=6.8cm,width=11.5cm]{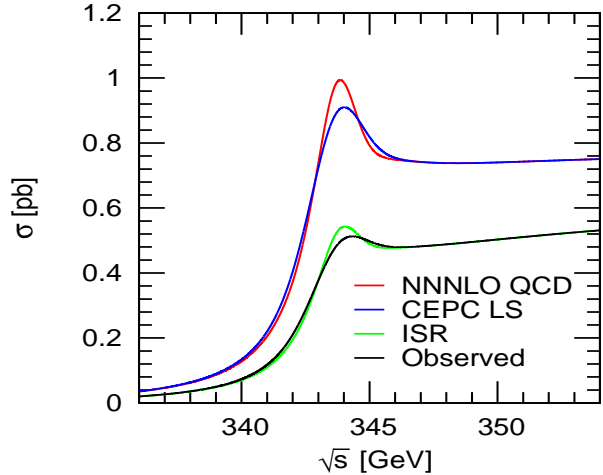} }
\caption{The red curve corresponds to the theoretical cross section at NNNLO QCD order obtained by the \texttt{QQbar$\_$threshold} code. The blue curve involves the luminosity spectrum(LS) correction at the CEPC. The green curve adds the impact from the initial state radiation(ISR) correction and the black curve is the experimental observed cross section.}
\label{various-cross-section}
\end{figure}

The experimental observed cross section is calculated by convoluting the theoretical cross section with the initial state radiation (ISR) factor and the luminosity spectrum (LS),
\begin{equation}\label{observed crossseciton}
\sigma_{t\bar{t}}^{obs}(\sqrt{s}) = \int_{0}^{\infty}\emph{d}\sqrt{s^{'}}\emph{G}(\sqrt{s^{'}},\sqrt{s})\cdot\int_{0}^{1}\emph{dx}\emph{F}(x,s^{'})\sigma_{t\bar{t}}^{th}(\sqrt{s^{'}}(1-x)),
\end{equation}
where \emph{G}($\sqrt{s^{'}},\sqrt{s}$) is the correction function due to large energy spread mainly caused by beamstrahlung and synchrotron radiation \cite{CEPC1,CEPC}, which  is usually described by a Gaussian function at the circular colliders \cite{CEPC2}, $\emph{F}(x,s^{'})$ is the initial state radiation factor \cite{FK-factor}, $\sigma_{t\bar{t}}^{th}(\sqrt{s^{'}}(1-x))$ is the theoretical cross section at NNNLO QCD order computed by \texttt{QQbar$\_$threshold} code. In Fig.\,\ref{various-cross-section}, the red curve is the purely QCD calculation of total cross section at NNNLO order. It can be seen clearly the so-called top pair bound state which locates just above the threshold energy point.  The blue and green curves are the total cross sections that are corrected by luminosity spectrum\,(LS) and ISR respectively. The black curve corresponds to the observed total cross section. We see that the ISR correction observably decreases the total cross section.

To simulate the top mass threshold scan experiment, we should assume an attempt top mass value as initial input parameter. The locations of the optimal energy points, which are determined by the simulation, may variate when the real top mass is different from our assumed value. However, the optimization method itself won't change and we are going to discuss this problem later after the simulation.

In our calculation, we adopt the PS mass scheme and our input parameters are set the same as in \cite{N3LO}. $\emph{$m_{t}$}^{\rm{PS}}(\mu_{f}$ = 20 GeV) = 171.5~GeV , $\Gamma_{t}$ = 1.33~GeV. Other input parameters are set as the default values in the \texttt{QQbar$\_$threshold} code. We approximately set energy spread as 0.1629\%, which is the designed energy spread at 240~GeV center of mass energy at CEPC \cite{CEPC,CEPC1,CEPC2} and we estimate it won't change much at $t\bar{t}$ threshold region.

\section{Top mass measurement scheme}
In this section, first, the experimental top mass measurements are simulated by using the software \texttt{Minuit} to perform the $\chi^2$ fittings. The number of optimal energy points and their locations are determined. Second, a substitutable theoretical analysis on statistical error matrix in one-point scheme is provided, which can calculate the statistical error directly even without any data-sampling and fitting. Finally, the luminosity dependence of the top mass statistical error is analysed and the accuracy of the top mass that can be achieved at the future collider CEPC is   discussed.

The crucial problem for a top mass threshold measurement optimization is to determine the number of energy points to be taken and their locations. Actually, relative studies~\cite{simulTop,CLIC,N3LO1} have already been made to simulate the threshold scan at future $e^{+}e^{-}$ colliders. In these researches, the data taking schemes are usually designed as 10 equal-distance energy points in a selected threshold nearby energy region with equal luminosity distributed on each energy point. It is quite easy to imagine that these 10 points definitely will not contribute equally to the fitted top mass due to their different sensitivity to the variations of the top mass and cross sections. In fact, it has been shown~\cite{scan-method} in similar tau lepton mass threshold scan case that the most efficient scheme is to concentrate all luminosity on the single optimal energy point in one free mass parameter fitting(Sometimes the background cross section and the selecting efficiency can also be taken as free parameters to be fitted and this is called the multi-parameter fitting, which we do not consider here). We have checked this result and find that the additional energy points far from the optimal energy point have negligible contribution to minimize the statistical error of the fitted top mass, and this indicates that they are completely unnecessary.

To find the location of the single optimal energy point, we perform several $\chi^2$ fittings in one energy point scheme and the single energy point ranging from 342~GeV to 346~GeV with a step of 0.1~GeV. The general $\chi^2$ function takes the form:
\begin{equation}
 \chi^2 = \sum^{n}_{i=1} \dfrac{[N_{i}-\mu_{i}(m_{t})]^{2}}{\mu_{i}(m_{t})},
\end{equation}
where $n$ is the number of energy points, and $n=1$ in the single energy point case,  $N_{i}$ is the number of top pair events which is simulated by Possion sampling according to the Poisson expectation value $\mu_{i}(m_{t})$ of the $i$th energy point. $\mu_{i}$ is given by
\begin{equation}\label{poisson-expectation}
\begin{aligned}
\mu_{i} &= [\epsilon_{sig}\cdot Br_{\emph{Wb}}\cdot\sigma_{t\bar{t}}^{obs}(\sqrt{s_{i}},m_{t})+\sigma_{BG}] \cdot \mathscr{L}_{i}\\
&\sim \mathscr{L}^{i}_{eff}\cdot Br_{\emph{Wb}}\cdot\sigma_{t\bar{t}}^{obs}(\sqrt{s_{i}},m_{t}),
\end{aligned}
\end{equation}
where $\epsilon_{sig}$ is the top pair selecting efficiency; $Br_{\emph{Wb}}$ is the  branching ratio for the decays of $\emph{t} \to \emph{W}^{+} b$ and $\emph{$\bar{t}$} \to \emph{W}^{-} \emph{$\bar{b}$}$, and we set $Br_{\emph{Wb}}$ = 1; $\sigma_{t\bar{t}}^{obs}(\sqrt{s_{i}},m_{t})$ can be obtained from Eq.\,(\ref{observed crossseciton}), $\sigma_{BG}$ is the background cross section, and $\mathscr{L}^{i}_{eff}=\mathscr{L}_{i}\cdot \epsilon_{sig}$ is the effective luminosity for the $i$th energy point. Most of the backgrounds can be reduced by the invariant mass cut as we discussed in section \ref{section2}. It can also be suppressed by other selection cuts \cite{ILC}. The interference between the resonant top pair decay and the single top decay process is also suppressed by \emph{$v^{2}$} \cite{BG_suppressed} with \emph{v} being the top quark velocity. Therefore, the backgrounds can safely be neglected in such a clean experiment.

Generally speaking, the observable cross section $\sigma_{t\bar{t}}^{obs}$ has two variable parameters, the top quark PS mass $m_t$ and the ECM $\sqrt{s}$. The change of the top quark width, as well as the strong coupling constant due to the variation of $m_t$ and $\sqrt{s}$ can be neglected. For convenience of the numerical calculations, the shape of the cross section can be approximately taken as stable and the change of $m_t$ can only cause a horizontal shift along the energy axis. Thus, the two variables $m_t$ and $\sqrt{s}$ can be reduced to a single one $\sqrt{s}$ - 2$\Delta \emph{m}_{t}$  as

\begin{equation}\label{equal}
\sigma_{t\bar{t}}^{obs}(\sqrt{s};m_{t})=\sigma_{t\bar{t}}^{obs}(\sqrt{s}-2\Delta m_{t},m_{t0}),
\end{equation}
where $\Delta \emph{m}_{t}$ = $\emph{m}_{t}$ - $\emph{m}_{t0}$, $\emph{m}_{t}$ is the top quark PS mass and $\emph{m}_{t0}=m_{t}^{\rm{PS}}(\mu_{f}=20~\mbox{GeV})=171.5~\mbox{GeV}$ is our initial input parameter.

Besides the $\chi^2$ fitting, the statistical error of the top mass can also be obtained from the error matrix analysis. The covariant matrix is described by
\begin{equation}\label{error-matrix}
V = \dfrac{\sigma(m_{t};\sqrt{s})}{\mathscr{L}_{eff}} \left[ \dfrac{\partial \sigma(m_{t};\sqrt{s})}{\partial m_{t}}\right]^{-2}.
\end{equation}
The statistical error is just the square root of the covariance matrix \cite{minuit},
\begin{equation}\label{statistical-error}
\begin{aligned}
\delta\emph{m}^{\rm{stat.}}_{t} =\sqrt{\frac{\sigma(m_{t};\emph{$\sqrt{s}$})}{\mathscr{L}_{eff}}}\left[ \dfrac{\partial \sigma(m_{t};\sqrt{s})}{\partial m_{t}}\right]^{-1}.
\end{aligned}
\end{equation}
So with Eq.\,(\ref{statistical-error}) we can calculate the statistical uncertainty directly.

\begin{figure}[htbp]
\centering
\centerline{ \includegraphics[height=6.8cm,width=11.5cm]{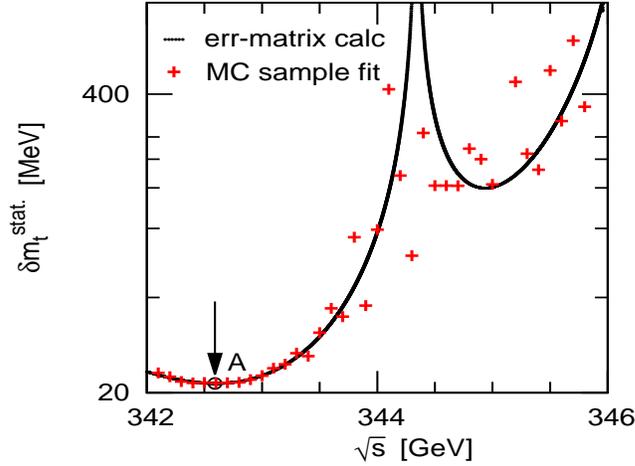} }
\caption{The relation between the statistical error $\delta m_{t}$ and the location of the single data-taken energy point in the $\sqrt{s}$-axis. The black curve is calculated by Eq.\,(\ref{statistical-error}) from the error matrix with fixed $\mathscr{L}_{eff}$ = 5 $fb^{-1}$, and the red cross dots correspond to the $\chi^2$ fitted results.}
\label{fit-calculate}
\end{figure}

\begin{figure}[htbp]
\centering
\centerline{ \includegraphics[height=6.8cm,width=11.5cm]{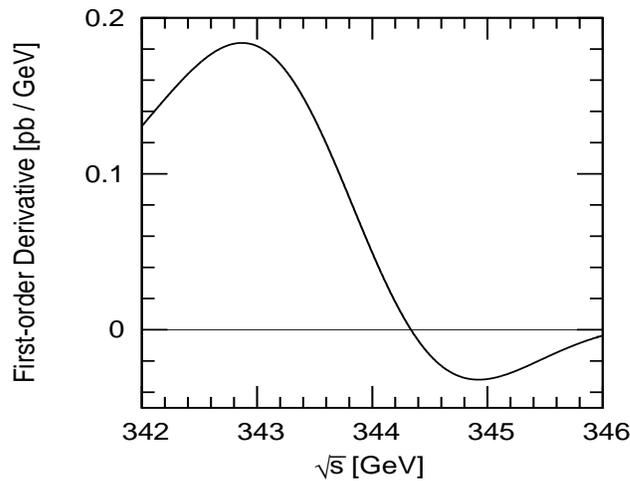} }
\caption{The first order derivative of the total cross section to the energy.}
\label{derivative}
\end{figure}

Fig.\ref{fit-calculate} shows the variation of the statistical error of the fitted top mass with different locations of the single energy point to be taken. The red crossed dots are our fitted results by \texttt{Minuit} with fixed $\mathscr{L}_{eff}$ = 5 $fb^{-1}$ at each energy point for each fitting. The black curve is the corresponding statistical uncertainties from the analytic calculation of error matrix by Eq.\,(\ref{statistical-error}). A point ``A"  at $\sqrt{s}\simeq342.6$ GeV is found to be the optimal energy point. It can been seen that this point locates near the largest slope but not exactly of the total cross section to the $\sqrt{s}$ in Fig.\ref{derivative}, as there is a $\sqrt{s}$ dependent term in front of the derivative shown in Eq.(\ref{statistical-error}).  From the figure, the statistical uncertainties from the analytic calculation agree well with that from the $\chi^2$ fitting in region $\sqrt{s} \in [342.0,344.0]$~GeV, but the consistency is not so good when the energy points approach to or above the threshold. The reason is due to the tiny value of the slope of the cross section here as shown in Fig.\,\ref{derivative}. Both the $\chi^2$ fitting by \texttt{Minuit} and the error matrix analysis do not have rapid convergence in this region and this indicates that it is a waste of luminosity to take energy point in this small slope region.

\begin{figure}[htbp]
\centering
\centerline{ \includegraphics[height=6.8cm,width=11.5cm]{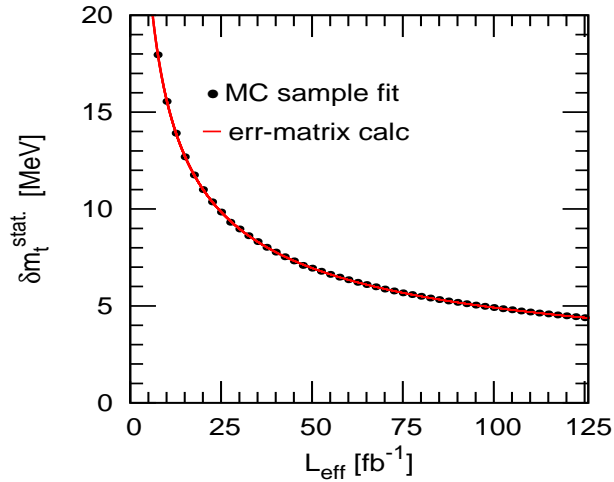} }
\caption{The correlation between the statistical error $\delta m_{t}^{stat.}$ and the $\mathscr{L}_{eff}$ in the optimal one-point $\sqrt{s} = 342.6$~GeV scheme. The black dots are $\chi^2$ fitted points, and the red curve is computed from error matrix analysis.}
\label{Chi2_error_luminosity_1point}

\end{figure}

Fig.\ref{Chi2_error_luminosity_1point} shows the decrease of the statistical uncertainty as the $\mathscr{L}_{eff}$ increase at the fixed optimal point $\sqrt{s}=342.6$~GeV. The red curve corresponds to statistical uncertainty from analytic calculation of error matrix and the black dots are fitted results by \texttt{Minuit}. Both of them coincide with each other. It can be seen that when $\mathscr{L}_{eff}=50fb^{-1}$, the statistical uncertainty is $\delta\emph{m}^{\rm{stat.}}_{t} =7~\mbox{MeV}$. Higher luminosity won't deduce to a significant decrease of the statistical error.

The theoretical uncertainty of the normalized total cross section at NNNLO QCD order is estimated at 3\% \cite{N3LO}. In our analysis, we assumed that the variation of top mass depends linearly on the total cross section, then the theoretical error of the top mass can be derived by the error transmission formula,
\begin{equation}\label{theoretical-error}
\delta m^{\rm{theory}}_{t} = \delta \sigma(\emph{$m_{t}$},\sqrt{s}) \cdot [\dfrac{\partial{\sigma(\emph{$m_{t}$},\sqrt{s})}}{\partial{m_{t}}}]^{-1}.
\end{equation}
Substituting the approximate formula $\partial\sigma(m_{t};\sqrt{s})/\partial m_{t}=2\partial\sigma(m_{t};\sqrt{s})/\partial\sqrt{s}$ into Eq.\,(\ref{theoretical-error}),
the top mass theoretical uncertainty is extracted to be $\pm$ 25.6~MeV which is significant larger than the statistical uncertainty. Similar result has also be presented in \cite{N3LO1}. For systematic uncertainty, simulation study shows it is expected to be about 10~MeV at FCC-ee~\cite{FCCee}. Without making careful analysis, which depends on the detailed information of the hardware, we expect here an equal value of the systematic error at CEPC. Thus, the total accuracy of the PS top mass that can be measured at CEPC is estimated as about $\delta m^{total}_{t}\sim30~\mbox{MeV}$, in which the statistical and systematic errors are comparable and the theoretical error is the dominate source.

Considering the real experiment, the situation is somewhat different from our simulation as the initial input top mass is unnecessarily be equal to the real top mass. The solution is circulating the fitting until to an acceptable accuracy. This means we initially put in an attempting top mass and find the corresponding single optimal energy point location, accumulate some events here, do the fitting and get a measured top mass. Then we take this experimental measured top mass as input parameter to determine the new location of the single optimal energy point, take data, and do the fitting once again. The circular can be stopped until the statistical uncertainty is suppressed to to an expectable accuracy. So the single optimal energy point does not mean we only take one energy point data in the whole experiment but one energy point in each fitting. The fitting itself can be made for many times and this circulation is the so called ``data-driven" method.

In order to compare the different points selection schemes, we also perform 10 points scheme as employed in the simulations~\cite{CLIC,N3LO1}. We take energy points from 340~GeV to 349~GeV by a step of 1~GeV and assign averagely $5~fb^{-1}$ effective integrated luminosity for each point. The total effective 50 $fb^{-1}$ integrated luminosity results to about 15~MeV statistical uncertainty, which analogous results holds in Refs.~\cite{CLIC,N3LO1}. For comparison, obviously the one point scheme leads to a sizeable improvement, about 50\% decrease of the top mass statistical uncertainty.
\section{The impact of accurate top mass on electroweak vacuum stability}
The sensitivity of the electroweak vacuum stability to the top mass is usually performed in the pole mass scheme, so we need to convert the PS mass into the pole mass. The relation between the PS top mass and the pole mass with corrections up to NNNLO QCD order takes the form \cite{mpole_mps},
\begin{equation}
\begin{aligned}
\emph{$m_{t}$}^{\rm{pole}} = &\emph{$m_{t}$}^{\rm{PS}}(\mu_{f}) +\dfrac{\mu_{f}C_{F}\alpha_{s}(\mu)}{\pi}[1 + \dfrac{\alpha_{s}(\mu)}{4\pi}(2\beta_{0}l_{1} + a_{1})
+ (\dfrac{\alpha_{s}(\mu)}{4\pi})^{2}(4\beta_{0}^{2}l_{2} + 2(2a_{1}\beta_{0} + \beta_{1})l_{1}\\ &+ a_{2}) + (\dfrac{\alpha_{s}(\mu)}{4\pi})^{3}(8\beta_{0}^{3}l_{3} + 4(3a_{1}\beta_{0}^{2} + \dfrac{5}{2}\beta_{0}\beta_{1})l_{2} +
2(3a_{2}\beta_{0} + 2a_{1}\beta_{1} + \beta_{2})l_{1} + a_{3} + \\&16\pi^{2}C_{A}^{3})],
\end{aligned}
\end{equation}
where $C_{F}$ = 4/3, $C_{A}$ = 3, $l_{1}$ = $\ln(\mu/\mu_{f})$ + 1, $l_{2} = \ln^{2}(\mu/\mu_{f}) + 2\ln(\mu/\mu_{f}) + 2$, $l_{3} = \ln^{3}(\mu/\mu_{f}) + 3\ln^{2}(\mu/\mu_{f}) + 6\ln(\mu/\mu_{f}) + 6$, $\mu_{f}$ is the subtraction scale and we set $\mu_{f}$ = 20~GeV for consistence with the setting $\emph{$m_{t}$}^{\rm{PS}}(\mu_{f}$ = 20~GeV) = 171.5~GeV, $\mu$ is the renormalization scale as we mentioned above and we set $\mu$ = 80 GeV, $\beta_{0}$, $\beta_{1}$, $\beta_{2}$ are the renormalization QCD $\beta$-functions calculated in \cite{betafunction}, and $a_{1}$, ${a_{2}}$, $a_{3}$ are constant coefficients related to the color factors and the number of light quarks, as given in \cite{a1a2beta1beta2constant,mpole_mps}.

The top pole mass reads
\begin{equation}
{m}_{t}^{\rm{pole}} = 173.294  \pm 0.007(\mbox{stat.}) \pm 0.026(\mbox{theory})\pm\mathcal{O}(0.2)(\mbox{ambiguity})~\mbox{GeV},
\end{equation}
where the three-loop strong running coupling \cite{threeloop_alphas} has be used. We'd like to point out there is a little different convention that the second and the third order coefficients of QCD $\beta$-functions in \cite{threeloop_alphas} are multiplied by a factor of 0.5, comparing with those in Ref.~\cite{betafunction}. Obviously, the uncertainty of the top pole mass is dominated by the intrinsic ambiguity which is estimated to be $\mathcal{O}$(200)~MeV \cite{ambiguity}. Both experimental efforts and high order theoretical calculations can not contribute to reduce this intrinsic uncertainty.

As far as the studies of vacuum stability at colliders concerned, the LHC could extract the Higgs boson mass with an accuracy of $\mathcal{O}$(200)~MeV \cite{pdg} and top quark pole mass with an accuracy of $\mathcal{O}(1)$~GeV \cite{LHC1}, as concluded in Ref.\,\cite{vacuum-stability}. The stable vacuum can be excluded at 98\% confidence level(C.L.) and only a small stable vacuum region is left in the [$m_{h}$,$m^{\rm{pole}}_{t}$] contour.  At future ILC, the top quark pole mass is estimated with an accuracy of 200~MeV  and uncertainties of Higgs boson is assumed to be below 50~MeV. A metastable vacuum in the Stand Model is expected at 95\% C.L.\cite{vacuum-stability-ILC}. At future CEPC, the Higgs boson mass can be extracted with an experimental accuracy of $\mathcal{O}$(10)~MeV \cite{higgs-CEPC}. Our research here shows at CEPC, the uncertainty of the top pole mass is also dominated by the irreducible ambiguity of the pole mass definition of $\mathcal{O}$(200)~MeV. Thus, the CEPC can have comparable or even better sensitivity to other $e^+e^-$ colliders to determine the vacuum stability in the Standard Model.

\section{Summary}
In this paper, the threshold scan of the top quark mass measurement experiment is simulated at future $e^+e^-$ collider near 350~GeV and the data taking strategy is optimized to minimize the statistical fluctuation of the top mass.
  The top pair production cross section adopted is up to NNNLO QCD level and the potential subtracted top mass is selected as it is free from the intrinsic ambiguity in the pole mass definition.
 The optimization shows the number of the energy point should be only one and it locates near the largest slope region of the cross section to the beam energy. Agreement result has been checked by the error matrix calculation of the statistical error.
 The so called data-driven method can be a best choice for the future real top mass measurement experiment which means the fitted top mass should be taken as a new input parameter to determine the next location of the optimal energy point and this circulative fittings can be stopped until the statistical error is minimized to an acceptable value. Our research has already shown the advantage of this optimized scheme.
 As data events are recorded at the most efficient single energy point, 7~MeV statistical uncertainty can be achieved within 50~$fb^{-1}$ effective integrated luminosity, which is about half of that comparing to the 10 average distributed points scheme used in~\cite{CLIC,N3LO1} with an equal total integrated luminosity.

The 3\% theoretical uncertainties for the normalized top pair production cross section caused by the renormalization scale variation leads to a 25.6~MeV theoretical uncertainties for the top mass.
 Although a hardware dependent systematic uncertainty analysis is still absent here, our simulation shows that the systematic error at the future CEPC is expected to be comparable to the statistical uncertainty at about $\mathcal{O}$(10)~MeV, the same as that at the FCC-ee~\cite{FCCee}.
 Actually, no matter whatever we can achieve to suppress the above error sources, the uncertainty of the top quark pole mass will still be dominated by the intrinsic ambiguity at about $\mathcal{O}$(200)~MeV.
 Even though, a top quark pole mass with $\mathcal{O}$(200)~MeV total uncertainty, together with the accurate Higgs mass measured at future $e^+e^-$ 250~GeV collider, is sufficient enough to make the final conclusion of the fate of the stability of the Standard Model electroweak vacuum.
 It is hard to believe that a metastable vacuum is just caused by coincidence and undiscovered new physics behind that is highly expected.

\begin{acknowledgements}
This work is supported by the National Science Foundation of China under the Grant No. 11405102 and the Fundamental Research Funds for the Central Universities of China under the Grant No. GK201603027.
\end{acknowledgements}

\bibliographystyle{apsrev}
%\bibliography{reference20170901}

\end{document}